\documentclass[a4paper]{article}

\usepackage{INTERSPEECH2022}
\usepackage{url}

\usepackage{graphicx}
\usepackage{float}

\usepackage{bm}
\usepackage{amssymb,amsmath,graphicx,bbm,color,booktabs}
\usepackage{amsopn}
\usepackage{multirow}

\newcommand{\vz}{\bm{z}}               

\newcommand{\R}{\mathbb{R}}

\newcommand{\reals}{\mathbb{R}}

\newcommand{\X}{\mathcal{X}}
\newcommand{\Z}{\mathcal{Z}}

\newcommand{\tableref}[1]{Table~\ref{#1}}

\renewcommand{\eqref}[1]{Eq.~(\ref{#1})}

\def \x{{\mathbf x}}

\newcommand{\figref}[1]{Fig.~\ref{#1}}
\newcommand{\mF}{\bm{F}} 

\title{Formant Estimation and Tracking using Probabilistic Heat-Maps}
\name{Yosi Shrem$^1$, Felix Kreuk$^2$, Joseph Keshet$^1$}
\address{
	$^1$Facullty of Electrical and Computer Engineering, Technion--Israel Institute of Technology, Israel\\
	$^2$Department of Computer Science, Bar-Ilan University, Ramat-Gan, Israel}
\email{ joseph.shrem@campus.technion.ac.il
, jkeshet@technion.ac.il}

\begin{document}
\maketitle

\begin{abstract}
Formants are the spectral maxima that result from acoustic resonances of the human vocal tract, and their accurate estimation is among the most fundamental speech processing problems. Recent work has been shown that those frequencies can accurately be estimated using deep learning techniques. However, when presented with a speech from a different domain than that in which they have been trained on, these methods exhibit a decline in performance, limiting their usage as generic tools.

The contribution of this paper is to propose a new network architecture that performs well on a variety of different speaker and speech domains. Our proposed model is composed of a shared encoder that gets as input a  spectrogram and outputs a domain-invariant representation. Then, multiple decoders further process this representation, each responsible for predicting a different formant while considering the lower formant predictions. An advantage of our model is that it is based on heatmaps that generate a probability distribution over formant predictions. Results suggest that our proposed model better represents the signal over various domains and leads to better formant frequency tracking and estimation.
\end{abstract}


 \section{Introduction}

Formants are resonances of the vocal tract. Typically there are 3 to 5 formant frequencies, roughly corresponding to each band of 1 kHz. As well as being essential for the perception of speech, they play an important role in speech coding, synthesis, and enhancement. A reliable estimation of these frequencies is essential in many phonological experiments in the fields of phonology, sociolinguistics, and bilingualism \cite{kathania2020study,kent2018static,hillenbrand1995acoustic, lee1999acoustics, vorperian2007vowel}.


Recently, it was proposed to use deep learning methods for the tasks of formant estimation and tracking \cite{dissen2016formant, dissen2019formant,dai2020formant}. The network was trained and tested on the Vocal Tract Resonance (VTR) database \cite{deng2006database} and achieved state-of-the-art results. However, as with traditional systems, experienced a dropoff in performance when the network was presented at test time with other recorded speech databases \cite{clopper2014effects, hillenbrand1995acoustic} due to the phenomenon of over-fitting to the speaker and speech domains represented in that database. 


Inspired by the tremendous advancement in object detection and pose estimation in the vision domain, we propose a similar yet modified modeling for the formant estimation based on probabilistic heatmaps. 
In the work of \cite{tompson2015efficient, wei2016convolutional,redmon2016you} heatmaps and probabilistic approaches were presented for estimating body parts and detecting objects rather than a direct coordinates prediction. We borrow ideas from those works to increase generalization in the task of formants estimation.

\begin{figure*}[h!]
	\centering   
	\advance\leftskip-1.05cm
	\includegraphics[trim={1.3cm 0 0 0},clip,height=5cm,width=20cm]{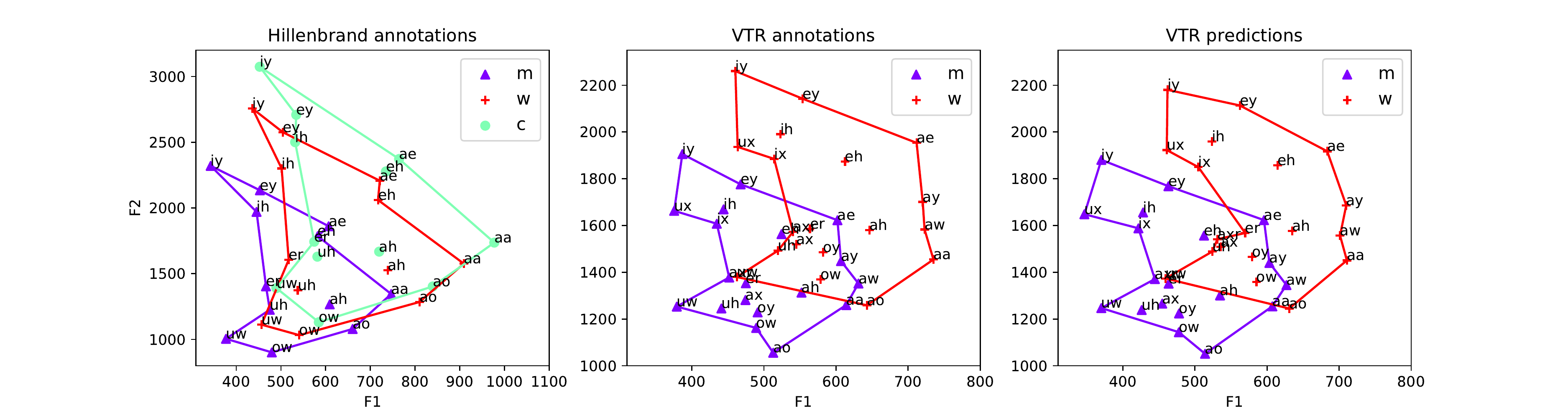}
	\caption{The average values for $F_1$ and $F_2$ for each vowel. Hillenbrand annotations (left), VTR annotations(middle), VTR predictions by our model (right). Groups are Men (m), Women (w) and Children (c).}
	\vspace{-12pt}
	\label{fig:vtr_hillenbrand}
\end{figure*}

Our goal is to train a single network that performs well on a variety of different speaker and speech domains. A straightforward approach is to use combined corpora consisting of multiple datasets instead of the VTR alone at the training phase.
However, when incorporating these databases into a single corpus, the results degraded for all domains of speech, including the VTR dataset. These datasets were collected for different research purposes, and there is significant variability among the datasets due to gender, age, dialects, recording conditions, etc.
A visualization of this phenomenon for the first two formants ($F_1$ and $F_2$) is depicted in the couple left sub-figures in \figref{fig:vtr_hillenbrand}. 
As one can see, there is a noticeable shift in the formants values between the gender and age groups. There is no direct transformation for the same group across datasets; the formants values among men recordings in VTR differ from those in Hillenbrand for the same vowel. Hence, direct optimization in the form of a regression task may not replicate the same performance if tested on samples from another dataset. 

This paper suggests a new modeling technique for formant tracking incorporating both the heatmap and the probabilistic approach. First, we use a convolutional neural network (CNN) model to generate a latent informative and explainable input representation without direct supervision. Then, multiple decoders are to predict formants as in a classification setup rather than traditional regression, which also improves generalization to new domains. Finally, an empirical evaluation of the proposed approach is presented on both in-domain examples as well as new datasets, and the generalization capability is discussed. 

\begin{figure*}[t]
	\centering   
	\includegraphics[height=3.3cm,width=17cm]{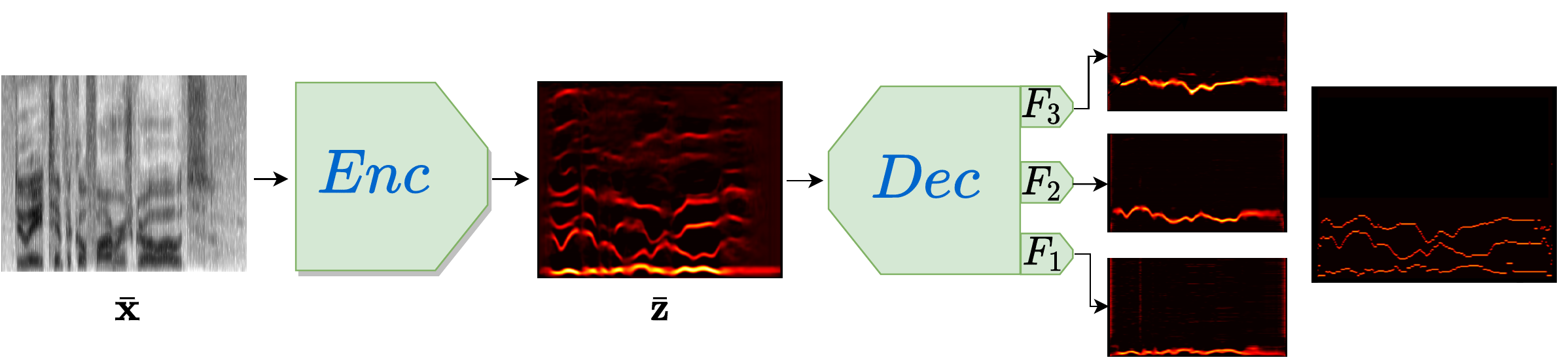}
	\vspace{-20pt}
	\caption{Our model. The spectrogram is encoded into a domain-invariant representation and  decoded into probabilistic heatmaps while preserving the original input dimensions. A heatmap aggregation is shown on the left, where the max value at each frame is taken as a prediction.}
	\label{fig:model}
\end{figure*}

\section{Problem settings}


The input to our model is a spectrogram mathematically represented as a sequence of $D$-dimensional real values, $\bar{\x}=(\x_1,\ldots,\x_T)$, where $\x_t \in \X \subseteq \R^D$ for $t \in [1, T]$, where $T$ is the number of frames in the input, and $D$ is the number of bins in the spectrogram. Associated with the spectrogram is an annotated sequence of $K$ formants $\bar{\mF}=(\mF^1,\ldots,\mF^T)$, where $\mF^t=(F^t_1,\ldots,F^t_K) \in \R^K$ is a vector of $K$ annotated formants of the $t$-th frame (we used $K=3$). Our goal is to predict the sequence $\bar{\mF}$.  

The proposed model is built from two components: an \emph{encoder} and multiple \emph{decoders}. The encoder generates a latent representation of the same size as the input spectrogram, namely $D\times T$. The encoder is denoted as a function $g: \X^T\to \Z^T$ from the domain of speech spectrogram to the domain of latent representations, $\Z \subseteq \reals^D$.  

The latent representation serves as input to a set of $K$ decoders that work jointly, where each decoder is responsible for estimating the conditional probability $\Pr(F_k | F_{k-1},\vz)$ of the $k$-th formant given the prediction of the lower formant and the latent frame representation $\vz$. At inference, the probability $\Pr(F_1|\vz)$  is estimated first, then the second formant probability is estimated given the first formant, $\Pr(F_2|F_1, \vz)$, and last, the third formant probability is estimated given the first two, $\Pr(F_3|F_2, F_1, \vz)$. Overall the formants at a given frame are predicted as the value $\mF^*=(F^*_1, F^*_2, F^*_3)$ that maximizes the conditional probability:
\vspace{-5pt}
\begin{multline}\label{eq:ideal_inference}
     \Pr(F_1, F_2, F_3 | \vz)  \\
     = \Pr(F_1|\vz) \Pr(F_2| F_1, \vz) \Pr(F_3|F_2,F_1,\vz)
     \vspace{-5pt}
\end{multline}

In our setting, the formant values are real numbers but quantized to $D$ bins. Then, the softmax function estimates the conditional probability of the quantized formant at each frame, and the median frequency at the associated bin is taken is predicted. The decoders are a set of functions $h_k$, $1\le k \le K$, where the decoder for $F_i$ receives as input the latent representation $\vz$, where the formants $F_j$ for $j < i$ have been masked.
We can consider the probability distribution over the formant values as \emph{heatmaps}. The heatmap is a graphical representation of the conditional probability space, showing the magnitude of a phenomenon based on the color of every area. A spectrogram is a heatmap visualization where the frequency is being squashed into bins, and the value of each bin indicates the strength of the signal in a specific frequency range.

\section{Proposed model}
In this section, we go over the model architecture and the optimization process in detail. At first, we describe the suggested model, components, and motivation. 


\subsection{Background }
A fundamental property of ``working in the wild'' is the ability to leverage information from a given data to new and unseen recordings. When splitting a given set of examples into train and test i.i.d, we rely on the assumption that both sets belong to an unknown yet particular distribution. Unfortunately, this assumption does not always hold while relating future samples as a new test set, i.e., new recordings may come from various acoustic environments or different speakers. 
Needless to say, such distribution shifts are critical and may frequently lead to severe performance degradation. In our case, the main issue limiting current methods is the fixation of a trained model to the frequency range where each formant was mainly seen during training. For example, as depicted in the middle subfigure in \figref{fig:vtr_hillenbrand}, the frequency of $F_1$ in a particular dataset speaker ranges between $350$ to $750$Hz; thus, a new sample with $F_1$ at $1,000$Hz can be presumed as out-of-distribution. Such phenomena could be due to the speaker's age or gender. In that case, the model behavior is undefined. In this paper, we address this issue and provide a robust formant tracking system consisting of two components - the first is invariant to the frequency range while the latter is.


Tracking formants can be broken down into two steps: (i) creating a shared representation, where information regarding all trackable formants is extracted using the encoder block; and (ii) predicting a probabilistic heatmap per formant ($F_1, F_2,$ and $F_3$) using the set of decoders $\{h_k\}_{k=1}^K$. By maintaining the spectrogram dimensions, we have a one-to-one mapping between a formant activation in the output heatmap and the corresponding frequency bin in the spectrogram. We begin by describing the shared representation encoder and then describe the set of decoders.

\subsection{Shared encoder}


As distribution shifts in unseen data are often presented as shifts of the spectra (e.g., an unseen speaker with a higher than usual pitch), our goal is to learn a representation that is robust to such changes. One way of incorporating such inductive bias into our encoder is to use 2D-convolutional layers on the spectrogram, which are translation-invariant to changes in the spectra. In other words, a formant should be detected even when it manifests in an unusual frequency band.

As in the encoder-decoder framework, supervision for the encoder output is not required. By being invariant to shifts in the spectra and receiving supervision in the form of gradients from the individual decoders, the encoder learns the general structure of formants in the spectrogram space, regardless of their prototypical frequency band. Upon closer inspection of the encoder output in \figref{fig:model}, we can see that it successfully captures additional formants in the given spectrogram (beyond $F_1, F_2, F_3$), even for untracked formants originating in higher frequency bands.

The encoder consists of $3\times 3$ kernels with residual connections and layered as follows: $1 - 16 - 32 - 64 - 128 - 128 - 64 - 32 - 1$, where the input channel dimension is one and the output as well. The height and width preserve the original dimension. Batch normalization and dropout layers are combined between the layers.  

 
\subsection{Multiple decoders}
Given the heatmap produced by the shared encoder, the task of the multiple decoders is to estimate the formants frequencies accurately by generating three formant-specific probabilistic heatmaps. At this stage, each decoder has to capture only the relevant information related to the associated formant while ignoring the rest of the intermediate heatmap. A masking function is required to prevent multiple heads from tracking the same formant. Thus, the decoding works sequentially by setting the input for each head as the shared representation where the output of previous heads is masked. As opposed to the encoder, the decoding phase consists of 1D-convolution. The convolved filter spans over the entire frequency axis, allowing it to detect in various optional bands. The output of each head is a probabilistic heatmap targeting a specific formant.

The decoder consists of three identical heads; each has a three-layer bottleneck with residual connection starting with $257 - 64 - 257$ where each kernel has a window of 3 frames, and the output shape of $257$ reconstructs the original spectrogram dimensions. Interestingly, in preliminary experiments, we observed that the bias term in these convolutional layers tends to converge to the prototypical location of the learned formant, leaving the kernel weights to be near zero (i.e., ignoring the input). To prevent this phenomenon, we remove the bias term in these layers. Batch normalization and dropout function are combined between the layers.

\section{Evaluation}

\vspace{-2pt}

\begin{table*}[h!]
	
	\renewcommand{\arraystretch}{1.1}
	\centering
	\footnotesize

	\caption{Formant tracking difference reported in mean absolute error on broad phone classes in VTR.} 
			\vspace{-5pt}
	\label{tab:tracking}  
	\resizebox{\linewidth}{!}{
		\begin{tabular}{l c c c|	c c c|c c c|c c c| c c c|c c c}
			\hline
			\hline
			&\multicolumn{3}{ c| }{Inter-labler} &\multicolumn{3}{ c| }{ Praat}  & \multicolumn{3}{ c| }{WaveSurfer}& \multicolumn{3}{ c| }{MSR}& \multicolumn{3}{ c| }{DeepFormant} & \multicolumn{3}{ c}{ Ours}  \\
			\cline{2-19}
			& $F_1$& $F_2$&$F_3$& $F_1$& $F_2$&$F_3$& $F_1$& $F_2$&$F_3$& $F_1$& $F_2$&$F_3$& $F_1$& $F_2$&$F_3$& $F_1$& $F_2$&$F_3$\\
			\cline{2-19}
			
			vowels              &$55$ & $69$ &$ 84$     &$130	$ &$230$ &$267$  &$70$ &$94$ &$154$   &$64$&$105$&$125 $ &$ \bf{54}$&$81$ &$112$
			&$57$ &$\bf{75}$&$\bf{95}  $    \\
			semivowels      &$ 68 $&$ 80 $&$ 103	  $&$136 	$&$295 $&$334  $&$89 $&$126 $&$222  $&$83$&$122$&$154   $&$\bf{67} $&$114 $&$168  $&$\bf{67} $&$\bf{90} $&$\bf{130} $\\
			nasal      &$ 75 $&$112 $&$ 106	$&$ 219	$&$409 $&$381 $
			&$ 96$&$229 $&$239   $&$67$&$\bf{120}$&$\bf{112}   $
			&$\bf{66}$&$175$&$151 $&$74 $&$130 $&$125$\\
			fricatives           &$ 91 $&$ 113 $&$ 125	$&$ 564 $&$593 $&$700 $
			&$  209$&$263 $&$439    $&$129$&$\bf{108}$&$\bf{131}  $&$131 $
			&$135 $&$159  $&$\bf{112} $&$120$&$140$\\
			affricatives       &$ 89 $&$ 118 $&$ 135	 $&$730  $&$515 $&$583   $  
			&$  292   $&$407 $&$ 390$&$  141$&$\bf{129}$&$149$&$  164 $
			&$162 $&$189 $&$\bf{140}  $&$138$&$\bf{145}$\\
			stops                &$ 91 $&$110 $&$116 	$&$258  $&$270 $&$351    $
			&$ 168    $&$210 $&$286 $&$   130$&$\bf{113}$&$\bf{119}$
			&$  131$&$135 $&$168    $&$\bf{117}$&$ 124  $&$130$\\
			
			\hline
			\hline
	\end{tabular} }
\vspace{-10pt}
\end{table*}

\subsection{Datasets}

In order to estimate a model's cross-datasets generalization performance, we used three corpora of manually annotated formants, each of which consists of different speakers and a recording setup.
The first dataset is the Vocal Tract Resonance (\textbf{VTR}) \cite{deng2006database}, a subset of the TIMIT dataset that has been split into train and test set containing  346 and 192 utterances, respectively. There are 173 speakers in the train set and 24 in the test set, presenting a diverse selection of speakers, dialects, gender, and phoneme. The first three formants($F_1$-$F_3$) were initially estimated using the MSR algorithm \cite{deng2004structured} and then were corrected manually. Worth mentioning that the fourth formant is also annotated; however, it has not been manually corrected and is based only on an automatic extraction. Hence we do not evaluate it in this paper.
The second dataset collected by \textbf{Clopper} and Tamati  \cite{clopper2014effects} consists 777 utterances produced by 20 female speakers in the age of 18-22. Each utterance contains a single word with a single vowel, and the $F_1$ and $F_2$ were annotated at the midpoint of the vowel.
The third dataset collected by \textbf{Hillenbrand} et al. \cite{hillenbrand1995acoustic} during a laboratory study consists of 1668 recordings produced by 45 men, 48 women, and 46 children (27 boys and 19 girls). The participants read a list of 12 single vowel words, and each recording was later split into 12 short utterances, one for each word.
The first three formants ($F_1$-$F_3$) were annotated at 4 points across the vowel- at 20\%,50\%, 80\% of the vowel duration and in additional time where the annotators have called a \emph{steady-point}.
For pre-processing we apply a pre-emphasis filter, $H(z) = 1-0.97z^{-1}$ , and the number of points for the FFT is $512$. Spectrograms were created with Torchaudio \cite{yang2021torchaudio}. The utterances sampling rate is  $16$ kHz; thus, each cell in both the spectrogram and heatmaps spans over approximately $31$ Hz.

\begin{table}[t!]
	
	\renewcommand{\arraystretch}{1.1}
	\centering
	\footnotesize
	\caption{Formant estimation error in Hz for multiple datasets and training techniques.} 
	\label{tab:clopper_hillenbrand_est}  
	\resizebox{\linewidth}{!}{
		\begin{tabular}{ l l c c c}
			\hline
			\hline
			Dataset& Method & $F_1$& $F_2$&$F_3$\\
			\hline
			\multirow{3}{*}{VTR} 
			&WaveSurfer					& $70$				& $96$ & $154$\\
			& DeepFormants	& $50$			&$86$& $104$ \\
			& \textbf{Ours}								& $\bf{39}$		&$\bf{30}$& $\bf{47}$ \\
			
			\hline
			\multirow{5}{*}{Hillenbrand} 
			&WaveSurfer					& $68$				& $190$ 		& $182$\\
			& DeepFormants (Train VTR)				& $71$				&$160$		  & $131$ \\
			& DeepFormants (Train All)				& $36$				&$100$		  & $116$ \\
			& \textbf{Ours}	 (Train VTR)							& $74$		& $150$    & $125$ \\
			& \textbf{Ours}	 (Train All)								& $\bf{26}$		& $\bf{78}$    & $\bf{82}$\\

			\hline
			\multirow{5}{*}{Clopper} 
			&WaveSurfer					& $128$			& $181$ 		& $-$\\
			& DeepFormants (Train VTR)				& $228$				&$168$		  & $-$ \\
			& DeepFormants (Train All)				& $103$		&$157$			& $-$ \\
			& \textbf{Ours}	 (Train VTR)							& $99$		& $147$    & $-$ \\
			& \textbf{Ours}	 (Train All)									& $\bf{49}$	& $\bf{64}$    & $-$ \\
			
			\hline
			\hline
	\end{tabular} }
\end{table}

\subsection{Augmentation \& hyperparameters}
Unfortunately, having numerous low-volume formants annotated datasets is a key challenge for training a deep learning model; hence, an artificial enlargement is required.
Data augmentations and manipulation are often used for increasing and enriching the training set. While traditional audio-related augmentation like input noising, random-cropping, and spectrogram reversal may boost a model performance, the modified instances are yet considered as in-distribution.
We applied an additional augmentation to simulate out-of-distribution samples synthetically, allowing the model to predict formants in a never seen range. Furthermore, we used a speed-up augmentation by a factor of two, directly counteracting the issue of a model focusing on popular frequency bands in the dataset. As a result, we maintained the ability to yield reliable annotations - the values were merely doubled, and the multi-head decoder was prevented from fixing the formants' values. Setting a non-integer speed-up factor resulted in inaccurate new labels: the originally annotated frame could have been undefined in the new representation, i.e., between adjacent time frames. Hence the augmentation must be tractable. In the following experiments, we optimized parameters using Adam \cite{kingma2014adam} with an initial learning rate of $1e-4$, annealed by a factor of 10 at epochs 300 and 600. In addition, we used label smoothing \cite{muller2019does} and randomly speeded up 20\% of the samples every epoch during training. 


						\vspace{-5pt}
\subsection{Formant estimation}
We begin by presenting our results for the estimation task and proceed with the tracking. The former demonstrates the generalization capabilities to a different domain, i.e., out-of-distribution.
We compare the suggested method with both analytically and machine-learning-based methods. Namely, (i) Praat \cite{boersma2011praat} and (ii) WaveSurfer \cite{sjolander2000wavesurfer}, widely used tools for phonetic research and speech analysis based on Linear Predictive Coding (LPC). We also report the results for the (iii) MSR algorithm \cite{deng2004structured} and DeepFormants \cite{dissen2019formant, dissen2016formant}. We followed the baseline setups as described in \cite{dissen2019formant}.

For the estimation task, which applies to vowels only, we averaged the frequencies for the entire vowel segment and reported the difference from the annotated test sample in \tableref{tab:clopper_hillenbrand_est}. Samples in the Clopper dataset are annotated only for $F_1$ and $F_2$; hence we leave blanks for the estimation difference in $F_3$. The results present two training methods: (i) training with VTR alone and estimating formants for Hillenbrand and Clopper, (ii) training with all three datasets and predicting for set-aside samples from each.
Since there is no predefined test set for Clopper and Hillenbrand, a third from each speakers group was chosen at random for the test set, leaving the rest for training.

As seen in the results, replicating the performance obtained from the trained samples to a new domain is a critical obstacle for machine-learning-based methods \cite{shrem2019dr,goldrick2021using}. There is a performance gap for DeepFormants and our model when we combine samples from the train set of Clopper and Hillenbrand during the optimization compared to train only with the VTR. The difference in $F_1$ estimation dropped by around 50\% for each model when we used the entire corpora during training. 
Our heatmap approach presented better performance than Wavesurfer and DeepFormants in both cases. Moreover, where only the VTR was used for training, the suggested method outperformed the previous system even where the latter was trained with samples from those datasets. 

To demonstrate the learned underlying structure, we computed the average annotated and predicted vowel frequencies in the VTR for each speakers group (men, women, children) and plotted them with the surrounding polygon in \figref{fig:vtr_hillenbrand}.

\subsection{Formant tracking}

This section addresses the VTR alone, being the only annotated frame-wised corpus. We consider tracking formants in VTR as an in-distribution task since utterances from the train and the test set share the same recording setup.



We compared our predictions with the annotations for each spoken frame and reported the mean absolute error in Hz for each phonetic class in \tableref{tab:tracking}. 
The authors of the VTR dataset \cite{deng2006database} reported the variation among the annotators for each formant and phonetic class as an anchor for evaluation.
When tested on the vowel and semivowel segments, the probabilistic approach presented here outperformed the previous approaches with a negligible imprecision in $F_1$ among vowels.
Specifically, the difference for the most indicative formant, i.e., $F_1$, is on par with the expected variation across human annotators or the most accurate among previous methods, i.e., 57 and 67 Hz difference in $F_1$ in vowels and semivowels compared to 55 and 68 Hz with human annotator, respectively. 
Results suggest a minor difference in $F_2$ vs. $F_3$ in the vowels phone class. However, it shows the most accurate measurements in comparison to previous automated systems.
The higher precision in the consonant class presented by MSR is highly likely due to the VTR annotation process; the initial formants were extracted using the MSR algorithm and manually corrected afterward.
In general, the higher the formant, the more difficult it is to track it, possibly due to its broad frequency bandwidth movement. An interesting phenomenon in our measurements is that the difference in $F_2$ is higher than $F_3$ among nasals - possibly a result of inaccurate annotating since most models reproduce the same behavior as the inter-labeler variance suggests.

A second assessment of the tracking performance involves analyzing the response time for a model to track a formant when switching from vowel to consonant (VC) and vice versa. At these transition points, the formant frequency shifts rapidly and has to be tracked again at the appropriate bandwidth. In line with previous study \cite{deng2006database,dissen2019formant,dissen2016formant}, we consider three frames before and after the transition as a six-frames window and calculate the mean absolute difference as reported in  \tableref{tab:CV_tiny}.

\begin{table}[t!]
	
	\renewcommand{\arraystretch}{1.1}
	\setlength{\tabcolsep}{4pt}
	\centering
	\footnotesize
	\caption{Formant tracking error in Hz for a six framed window with a transition from consonant to wel and vice versa.} 
			\vspace{-5pt}
	\label{tab:CV_tiny}  
		\begin{tabular}{cc c c|c c c| c c c}
			\hline
			\hline
			& \multicolumn{3}{ c| }{MSR}& \multicolumn{3}{ c| }{DeepFormant} & \multicolumn{3}{ c}{Ours}  \\
			\cline{2-10}
			& $F_1$& $F_2$&$F_3$& $F_1$& $F_2$&$F_3$& $F_1$& $F_2$&$F_3$\\
			\cline{2-10}
			
			CV transitions         &$106$&$101$&$119   	 $&$ 110$&$142 $&$166         $&$\bf{91} $&$\bf{85}$&$\bf{97}$      \\
			VC transitions 				  &$\bf{48} $&$92 $&$120 	 $&$53$&$80$&$111        		$&$59 $&$\bf{62} $&$\bf{77}$     \\
			%
			\hline
			\hline
			
	\end{tabular} 
\end{table}

\section{Conclusions}
\vspace{-5pt}
In this paper, we presented a probabilistic approach for the task of format tracking and estimation based on heatmaps classification rather than regression as in previous models. Focusing on generalization, we introduced novel modeling that can generate predictions with superior accuracy for both in-domain and out-of-distribution samples. In future work, we would like to further explore the generalization effect by combining additional augmentation techniques to simulate out-of-distribution samples. In addition, we believe the generated shared heatmap representation captures the acoustic structure of the speaker and can be used in a variety of speech-related tasks.
The code is publicly available at \textit{https://github.com/MLSpeech/FormantsTracker}.

\bibliographystyle{IEEEtran}
\bibliography{refs}

\end{document}